\documentclass[twocolumn]{pasj00}
\draft
\usepackage[dvips]{color}
\usepackage{multicol}

\begin{document}
\SetRunningHead{N. Narita et al.}{A Retrograde Planet HAT-P-7b}
\Received{2008/08/05}%{yyyy/mm/dd}
\Accepted{2008/08/27}%{yyyy/mm/dd}

\title{First Evidence of a Retrograde Orbit of
Transiting Exoplanet HAT-P-7b$^*$}

%%% begin:list of authors

\author{
Norio \textsc{Narita},\altaffilmark{1}
Bun'ei \textsc{Sato},\altaffilmark{2}
Teruyuki \textsc{Hirano},\altaffilmark{3}
Motohide \textsc{Tamura}\altaffilmark{1}
}

\altaffiltext{1}{
National Astronomical Observatory of Japan, 2-21-1 Osawa,
Mitaka, Tokyo, 181-8588, Japan
}
\email{norio.narita@nao.ac.jp}

\altaffiltext{2}{
Global Edge Institute, Tokyo Institute of Technology,
2-12-1 Ookayama, Meguro, Tokyo, 152-8550, Japan
}

\altaffiltext{3}{
Department of Physics, The University of Tokyo, Tokyo, 113-0033, Japan
}

%% `\KeyWords{}' always has to be placed before `\maketitle'.
%Do NOT move this preamble from here!

\maketitle

\begin{abstract}
  We present the first evidence of a retrograde orbit of the transiting
  exoplanet HAT-P-7b. The discovery is based on a measurement
  of the Rossiter-McLaughlin effect with the Subaru HDS
  during a transit of HAT-P-7b, which occurred
  on UT~2008~May~30.
  Our best-fit model shows that the spin-orbit
  alignment angle of this planet is $\lambda = -132.6^{\circ}$
  $^{+10.5^{\circ}}_{-16.3^{\circ}}$.
  The existence of such a retrograde planet have been predicted by 
  recent planetary migration models considering planet-planet
  scattering processes or the Kozai migration. Our finding provides
  an important milestone that supports such dynamic migration theories.
\end{abstract}
\footnotetext[*]{Based on data collected at Subaru Telescope,
which is operated by the National Astronomical Observatory of Japan.}

\section{Introduction}

One of the surprising properties of extrasolar planets
is their distributions around their host stars.
Since many Jovian planets have been found in the vicinity
(far inside the snow line) of their host stars,
numbers of theoretical models have been studied to
explain inward planetary migration.
Recently understanding of planetary migration mechanisms
has rapidly progressed through observations of
the Rossiter-McLaughlin effect (hereafter
the RM effect: \cite{1924ApJ....60...15R},
\cite{1924ApJ....60...22M}) in transiting
exoplanetary systems.
The RM effect is an apparent radial velocity anomaly during
planetary transits.
By measuring this effect, one can learn the sky-projected angle
between the stellar spin axis and the planetary orbital axis,
denoted by $\lambda$ (see \cite{2005ApJ...622.1118O, 2007ApJ...655..550G}
for theoretical discussion).

So far, spin-orbit alignment angles of about 15 transiting
planets have been measured (\cite{2009ApJ...696.1230F},
and references therein).
Among those RM targets, significant spin-orbit misalignments
have been reported for 3 transiting planets:
XO-3b \citep{2008A&A...488..763H, 2009ApJ...700..302W},
HD80606b (\cite{2009A&A...498L...5M};
\cite{2009A&A...502..695P, 2009arXiv0907.5205W}),
and WASP-14b \citep{2009arXiv0907.5204J}.
These misaligned planets are considered to have migrated
through planet-planet scattering processes
(e.g., \cite{1996Sci...274..954R, 2002Icar..156..570M, 
2008ApJ...678..498N, 2008ApJ...686..580C})
or Kozai cycles with tidal evolution
(e.g., \cite{2003ApJ...589..605W, 2005ApJ...627.1001T, 
2007ApJ...669.1298F, 2007ApJ...670..820W}),
rather than the standard Type II migration (e.g.,
\cite{1985prpl.conf..981L, 1996Natur.380..606L, 2004ApJ...616..567I}).

The existence of such misaligned planets has demonstrated
validity of the planetary migration models considering
planet-planet scattering or the Kozai migration.
On the other hand, such planetary migration models also predict
significant populations of ``retrograde'' planets.
Thus discoveries of retrograde planets would be an important
milestone for confirming the predictions of recent
planetary migration models, and intrinsically interesting.

In this letter, we report the first evidence of such
a retrograde planet in the transiting exoplanetary system HAT-P-7.
Section~2 summarizes the target and our Subaru observations,
and section~3 describes the analysis procedures for the RM effect.
Section~4 presents results and discussion for the derived
system parameters.
Finally, section~5 summarizes the main findings of this letter.

\section{Target and Subaru Observations}

HAT-P-7 is an F6 star at a distance of 320 pc
hosting a very hot Jupiter
(\cite{2008ApJ...680.1450P}; hereafter P08).
Among transiting-planet host stars,
F type stars are interesting RM targets
because these stars often have a large stellar rotational velocity,
which facilitates measurements of the RM effect.
However, the rotational velocity of HAT-P-7 is
$V \sin I_s = 3.8$~km~s$^{-1}$ (P08), which is unusually
slower than expected for an F6 type star.
Nevertheless, this system is favorable for the RM observations,
since the star is relatively bright ($V=10.5$) and
the expected amplitude of the RM effect
($V \sin I_s (R_p/R_s)^2 \sim 20$~m~s$^{-1}$)
is sufficiently detactable with the Subaru telescope.

We observed a full transit of HAT-P-7b with the High Dispersion
Spectrograph (HDS: \cite{2002PASJ...54..855N}) on the Subaru 8.2m
telescope on UT 2008 May 30.
We employed the standard I2a set-up of the HDS, covering
the wavelength range 4940~\AA\ $< \lambda <$ 6180~\AA\, and used the
Iodine gas absorption cell for radial velocity measurements.
The slit width of $0\farcs6$ yielded a spectral resolution of
$\sim$60000. The seeing on that night was around $0\farcs6$.
The exposure time for radial velocity measurements
was 6-8 minutes, yielding a typical signal-to-noise ratio (SNR)
of approximately 120 per pixel.
We processed the observed frames with standard IRAF\footnote{The Image
  Reduction and Analysis Facility (IRAF) is distributed by the U.S.\
  National Optical Astronomy Observatories, which are operated by the
  Association of Universities for Research in Astronomy, Inc., under
  cooperative agreement with the National Science Foundation.}
procedures and extracted one-dimensional spectra.
We computed relative radial velocities following the algorithm of
\citet{1996PASP..108..500B} and \citet{2002PASJ...54..873S}, as
described in \citet{2007PASJ...59..763N}.  We estimated the internal
error of each radial velocity as the scatter in the radial-velocity
solutions among $\sim$4~\AA~segments of the spectrum. The typical
internal error was $\sim$5~m~s$^{-1}$.
The radial velocities and uncertainties are summarized in table~1.

\section{Analyses}

We model the RM effect of HAT-P-7 following the
procedure of \citet{2005ApJ...631.1215W}, as described in
\citet{2009arXiv0905.4727N} and Hirano~et~al.~in~prep.
We start with a synthetic template spectrum, which matches for
the stellar property of HAT-P-7 described in P08,
using a synthetic model by \citet{2005A&A...443..735C}.
To model the disk-integrated spectrum of HAT-P-7,
we apply a rotational broadening kernel of $V \sin I_s =
3.8$~km~s$^{-1}$ and assume limb-darkening parameters
for the spectroscopic band as $u_1 = 0.45$ and
$u_2 = 0.31$, based on a model by \citet{2004A&A...428.1001C}.
We then subtract a scaled copy of the original unbroadened
spectrum with a velocity shift to simulate spectra during a transit.
We create numbers of such simulated spectra
using different values of the scaling factor $f$ and the
velocity shift $v_p$, and compute the apparent radial velocity of
each spectrum.  We thereby determine an empirical formula that
describes the radial velocity anomaly $\Delta v$ in HAT-P-7
due to the RM effect, and find
\begin{equation}
\Delta v = - f v_p \left[1.444 - 0.623
\left( \frac{v_p}{V \sin I_s} \right)^2 \right].
\end{equation}

For radial velocity fitting, including the Keplerian motion and
the RM effect,
we adopt stellar and planetary parameters based on P08 as follows;
the stellar mass $M_s = 1.47$ [$M_{\odot}$],
the stellar radius $R_s = 1.84$ [$R_{\odot}$],
the radius ratio $R_p/R_s = 0.0763$,
the orbital inclination $i = 85.7^{\circ}$,
and the semi-major axis in units of the stellar radius
$a / R_s = 4.35$.
We assess possible systematic errors due to uncertainties
in the fixed parameters in section~4.
We also include a stellar jitter
of $3.8$~m~s$^{-1}$ for
the P08 Keck data
as systematic errors of radial velocities by quadrature sum.
It enforces the ratio of $\chi^2$ contribution and
the degree of freedom for the Keck data to be unity.
We do not include additional radial velocity errors
for the Subaru data, because we find the ratio for the Subaru dataset
is already smaller than unity (as described in section~4).

In addition, we adopt the transit ephemeris $T_c = 2453790.2593$ [HJD]
and the orbital period $P = 2.2047299$~days based on P08.
Note that this ephemeris has an uncertainty of 3 minutes for the
observed transit; however the uncertainty is well within our
time-resolution (exposure time of 6-8 minutes and readout time
of 1 minute) and is negligible for our purpose.
The adopted parameters above are summarized in table~2.

Our model has 3 free parameters describing the HAT-P-7 system:
the radial velocity semi-amplitude $K$,
the sky-projected stellar rotational velocity $V \sin I_s$,
and the sky-projected angle between the stellar spin axis and
the planetary orbital axis $\lambda$.
We fix the eccentricity $e$ to zero,
and the argument of periastron $\omega$ is not considered.
Finally we add two offset velocity parameters for respective radial
velocity datasets
($v_1$: our Subaru dataset, $v_2$: the Keck dataset in P08).

We then calculate the $\chi^2$ statistic (hereafter ``main-case'')
\begin{eqnarray}
\chi^2 &=& \sum_i \left[ \frac{v_{i,{\rm obs}}-v_{i,{\rm calc}}}
{\sigma_{i}} \right]^2,
\end{eqnarray}
where
$v_{i, {\rm obs}}$ and $\sigma_{i}$ are observed radial velocities
and uncertainties,
and $v_{i, {\rm calc}}$ are radial velocity values calculated based
on a Keplerian motion and on the empirical RM formula given above.

We determine optimal orbital parameters by minimizing the $\chi^2$
statistic using the AMOEBA algorithm \citep{1992nrca.book.....P}.
We estimate 1$\sigma$ uncertainty of each free parameter based on
the criterion $\Delta \chi^2 = 1.0$ when a parameter
is stepped away from the best-fit value and
the other parameters are re-optimized.

We also fit the radial velocities using another statistic function
for reference (hereafter ``test-case''),
\begin{eqnarray}
\chi^2 &=& \sum_i \left[ \frac{v_{i,{\rm obs}}-v_{i,{\rm calc}}}
{\sigma_{i}} \right]^2
+ \left[ \frac{V \sin I_s - 3.8}{0.5} \right]^2.
\end{eqnarray}
The last term is {\it a priori} constraint for $V \sin I_s$
to match the independent spectroscopic analysis by P08.

\section{Results and Discussion}

Figure~1 shows observed radial velocities and the best-fit model curve
for the main-case.
Figure~2 illustrates the RM effect of HAT-P-7b
with the best-fit model
and also shows a comparison with the case of $\lambda=0^{\circ}$ and
$V \sin I_s = 3.8$~km~s$^{-1}$ model.
The upper panel of figure~3 plots a $\chi^2$ contour
in ($\lambda, V \sin I_s$) space.
As a result, we find the key parameter
$\lambda = -132.6^{\circ} \, (+10.5^{\circ}, -16.3^{\circ})$,
implying a retrograde orbit of HAT-P-7b.
The stellar rotational velocity is
$V \sin I_s = 2.3 \,\, (+0.6, -0.5)$~km~s$^{-1}$,
which is marginally consistent with the P08 spectroscopic result
($V \sin I_s = 3.8 \pm 0.5 $~km~s$^{-1}$).
Residuals from the best-fit model indicate rms of 4.14~m~s$^{-1}$
for the Subaru dataset and 4.09~m~s$^{-1}$ for the P08 Keck dataset.
The rms of the Subaru residuals is well within our
internal radial velocity errors, and that of the Keck residuals
is in good agreement with the assumed jitter level of 3.8~m~s$^{-1}$.
One may wonder that a smaller $V \sin I_s$ allowed in the main-case
would weaken the detection-significance of the RM effect.
However, since $V \sin I_s = 0$~km~s$^{-1}$ is excluded
by $\Delta \chi^2 = 36.49$, there is very
little chance that a true $V \sin I_s$ is actually nearly zero and
a spin-orbit alignment angle $\lambda$ is very small.
The lower panel of figure~3 plots a similar $\chi^2$ contour
but for the test-case. In this case, we find
$\lambda = -122.5^{\circ} \, (+6.4^{\circ}, -7.7^{\circ})$
and $V \sin I_s = 3.1 \pm 0.4$~km~s$^{-1}$.
Thus our two results (main and test cases) are well consistent
with each other.
In addition, we test the fitting with
the eccentricity $e$ and the argument of periastron $\omega$
as free parameters. As a result, we do not find any significant
(nonzero) eccentricity for this planet.
The best-fit parameters and uncertainties are summarized in table~3.

In the above analyses, we fixed several parameters
as summarized in table~2, which were based on P08 and
\citet{2004A&A...428.1001C}.
In order to estimate the level of possible systematic errors,
we retry the fitting for following four cases;
(1) $a / R_s = 4.63,\,\, i = 89.2^{\circ}$
(corresponding to 1$\sigma$ lower limit of the impact parameter in P08);
(2) $a / R_s = 3.97,\,\, i = 82.6^{\circ}$
(corresponding to 1$\sigma$ upper limit of the impact parameter in P08);
(3) $u_1 = 0.65$ (a greater limb-darkening case); and
(4) $u_1 = 0.25$ (a smaller limb-darkening case).

Consequently, we find that respective results for
$\lambda$ and $V \sin I_s$ are;
(1) $\lambda = -151.8^{\circ} \, (+11.1^{\circ}, -12.8^{\circ})$
and $V \sin I_s = 2.1 \pm 0.4$~km~s$^{-1}$;
(2) $\lambda = -99.5^{\circ} \, (+3.0^{\circ}, -6.2^{\circ})$
and $V \sin I_s = 7.6 \pm 2.9$~km~s$^{-1}$;
(3) $\lambda = -135.9^{\circ} \, (+11.2^{\circ}, -16.7^{\circ})$
and $V \sin I_s = 2.2 \pm 0.5$~km~s$^{-1}$; and
(4) $\lambda = -129.8^{\circ} \, (+9.9^{\circ}, -15.6^{\circ})$
and $V \sin I_s = 2.4 \pm 0.6$~km~s$^{-1}$.
Thus $\lambda = -164.6^{\circ}$ -- $-96.5^{\circ}$
and $V \sin I_s = 1.7$ -- $10.5$~km~s$^{-1}$
can be still probable
if the uncertainties for fixed parameters (especially for
the impact parameter) are taken into account.
These systematic errors would be significantly reduced
when the Kepler photometric data for HAT-P-7
are available \citep{2009Sci...325..709B}.

The derived value of $\lambda$ seems
to indicate a retrograde orbit by itself.
However, since the true spin-orbit angle $|\Psi|$ also
depends on the inclination of the stellar spin axis $I_s$,
$|\Psi|$ is not necessarily greater than $90^{\circ}$
(corresponding to a retrograde orbit)
even if $\lambda = -132.6^{\circ}$.
Thus one might wonder whether the planet HAT-P-7b is
statistically in a retrograde orbit.
We can roughly estimate the probability using
the relation of spherical geometry,
\begin{eqnarray}
\cos |\Psi| = \cos I_s \cos i + \sin I_s \sin i \cos \lambda.
\end{eqnarray}
Note that $I_s$ ranges from $0^{\circ}$ to $180^{\circ}$.
We compute the true spin-orbit angle $|\Psi|$ of HAT-P-7b
by substituting the observed values of $i$ and $\lambda$
into the relation.
We adopt $i = 85.7^{\circ}$ (P08) and
test three representative cases for $\lambda$
($-164.6^{\circ}, -132.6^{\circ}, \,\,\rm{and}\, -96.5^{\circ}$).
Assuming an uniform distribution for $\cos I_s$ within
the range of value, the probabilities of a retrograde orbit
($|\Psi| > 90^{\circ}$) are 99.85\% ($\lambda=-164.6^{\circ}$),
99.70\% ($\lambda=-132.6^{\circ}$), and
91.65\% ($\lambda=-96.5^{\circ}$), respectively.
We note that $|\Psi|$ is always larger than $85.7^{\circ}$
(the adopted value of $i$, in the case of $I_s = 0^{\circ}$).
Those estimates favor a retrograde orbit of HAT-P-7b.

On the other hand, it is important to point out that the stellar
rotational velocity $V \sin I_s = 3.8$~km~s$^{-1}$ determined by
the spectroscopic analysis (P08) is exceptionally slow for an F6 star.
For example, HAT-P-2 and TrES-4, which are
other known planetary host stars with similar spectral type,
have larger stellar rotational velocities:
$V \sin I_s = 19.8$~km~s$^{-1}$
(HAT-P-2, \cite{2007ApJ...670..826B} and also confirmed
from the RM effect by \cite{2007ApJ...665L.167W}) and
$V \sin I_s = 8.5$~km~s$^{-1}$
(TrES-4, \cite{2009ApJ...691.1145S} and also confirmed
from the RM effect by Narita et al. in prep.).
The small $V \sin I_s$ may suggest a smaller inclination angle
of the stellar spin axis.
In that case, it is highly possible that the planet is
in a nearly polar retrograde orbit.

We note that too small $I_s$ can be constrained by the facts
that a faster stellar rotation of HAT-P-7 than 500~km~s$^{-1}$
would be physically unlikely due to stellar break-up,
and that a faster rotation than 100~km~s$^{-1}$ would
be empirically unlikely for a F6 star \citep{2005oasp.book.....G}.
Translating the constraints into $\cos I_s$, we find
a probability of such unrealistic cases
is only $\sim$0.03\%, which has very little impact on
the probability estimations for a retrograde orbit.
In any case, it would be important to directly constrain $I_s$
by other observational methods (e.g., \cite{2008AJ....135...68H}
or asteroseismology with the Kepler) in order to estimate a
true spin-orbit angle of HAT-P-7b.

We previously experienced a false positive
of a spin-orbit misalignment in HD17156b due to lower
precision radial velocity data
(\cite{2008PASJ...60L...1N, 2009arXiv0905.4727N}).
The problem for the HD17156 case \citep{2008PASJ...60L...1N}
was that radial velocity uncertainties were comparable with
predicted RM amplitude and also due to a poor number
of radial velocity samples.
Based on the lesson, we estimate the significance of
our RM detection using the equation (26) in
\citet{2007ApJ...655..550G}.
As a result, the SNR of our RM detection is over 10,
and thus we conclude to have obtained radial velocities
of a sufficient number and precision
to model the RM effect of HAT-P-7b.

Nevertheless, we finally note that we should care about possible
systematic errors in $\lambda$.
Since the RM amplitude of HAT-P-7b is only $\sim15$~m~s$^{-1}$,
any systematic shift as much as several m~s$^{-1}$
due to stellar jitter or other reasons
at ingress or egress phase would
cause a large systematic difference in $\lambda$.
Thus further radial velocity measurements for this interesting
system are desired to confirm a retrograde orbit of HAT-P-7b
more decisively.

\section{Summary}

We observed a full transit of HAT-P-7b with the Subaru 8.2m telescope
on UT 2008 May 30, and measured the RM effect of this planet.
Based on the RM modeling, we discovered the first evidence of
a retrograde orbit of HAT-P-7b.
This is the first discovery of a retrograde extrasolar planet.
The existence of such planets have been indeed predicted
in some recent planetary migration models considering
planet-planet scattering and/or the Kozai migration
(e.g., \cite{2007ApJ...669.1298F, 2007ApJ...670..820W};
\cite{2008ApJ...678..498N, 2008ApJ...686..580C}).
In addition, it is interesting to point out that HAT-P-7b
is the first spin-orbit misaligned planet having
no significant eccentricity.
This discovery may suggest that other hot Jupiters in circular orbits
also have significant spin-orbit misalignments or even retrograde
orbits.
Thus further RM observations for transiting planets,
including not only eccentric or binary system planets
but also close-in circular planets,
would be encouraged in order to understand populations
of aligned/misaligned/retrograde planets.

\textit{Note added after submission ---}
\citet{2009arXiv0908.1672W} report an independent evidence for
a retrograde orbit of HAT-P-7b, based on independent Subaru HDS
observations conducted in 2009 June/July.

\bigskip
We acknowledge the invaluable support for our Subaru observations
by Akito Tajitsu, a support scientist for the Subaru HDS.
We are grateful to Yasushi Suto, Ed Turner, Wako Aoki, and Toru Yamada
for helpful discussions on this study;
Josh Winn and his colleagues for sharing
their information in advance of publication.
This paper is based on data
collected at Subaru Telescope, which is operated by the National
Astronomical Observatory of Japan.  The data analysis was in part
carried out on common use data analysis computer system at the Astronomy
Data Center, ADC, of the National Astronomical Observatory of Japan.
N.N. is supported by a Japan Society for Promotion of Science (JSPS)
Fellowship for Research (PD: 20-8141).
We wish to acknowledge the very significant cultural role
and reverence that the summit of Mauna Kea has always had within
the indigenous Hawaiian community.

%%%%%%%%%%%%%%%%%%%%%%%%%%%%%%%%%%%%%%%%%%%%%%%%%%%%%%%%%%%%%%%%%%%%%%

%%%%%%%%%%%%%%%%%%%%%%%%%%%%%%%%%%%%%%%%%%%%%%%%%%%%%%%%%%%%%%%%%%%%%%

\clearpage
%%%%%%%%%%%%%%%%%%%%%%%%%%%%%%%%%%%%%%%%%%%%%%%%%%%%%%%%%%%%%%%%%%%%%%
\begin{figure*}[pthb]
 \begin{center}
  \FigureFile(160mm,160mm){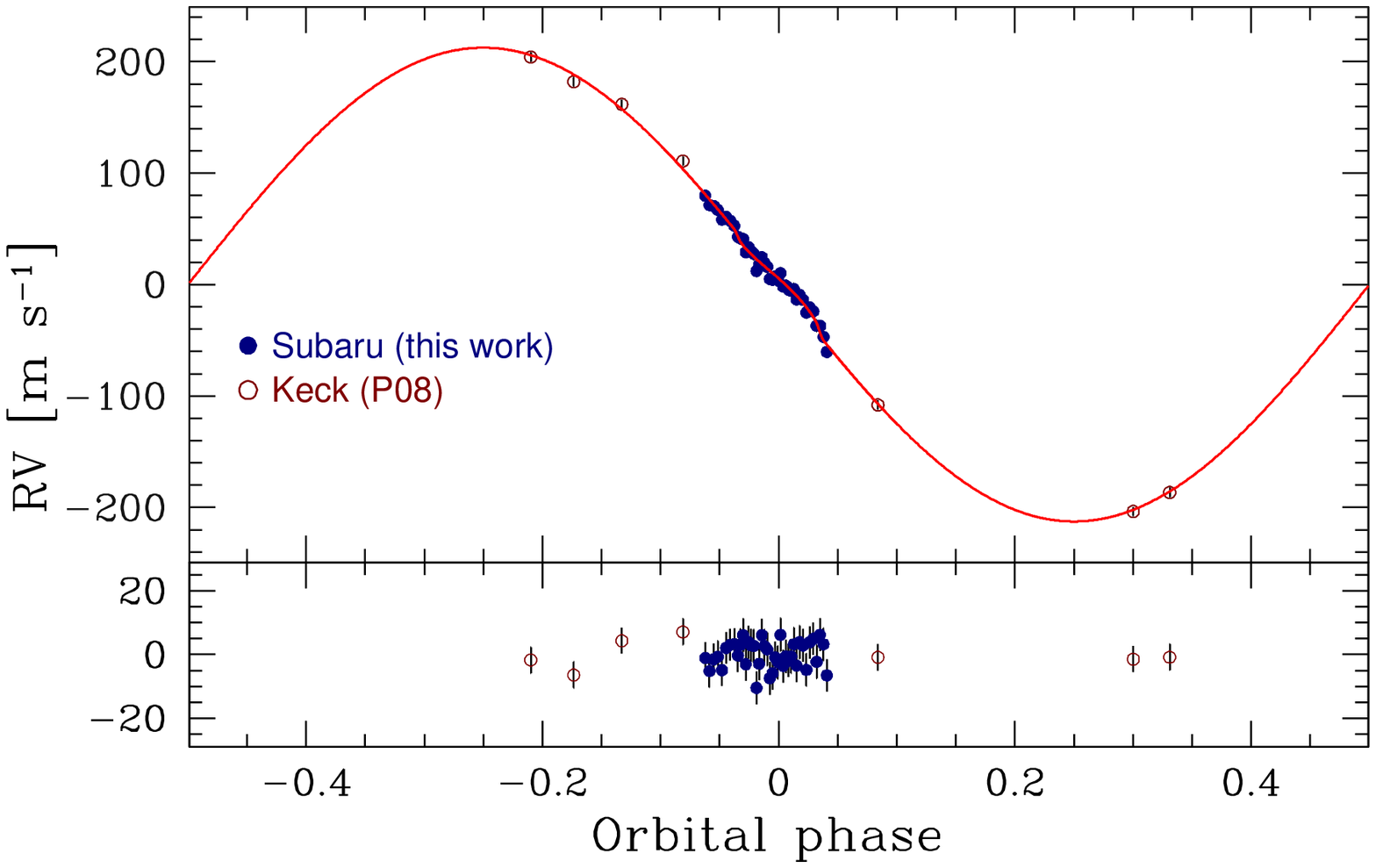}
  \FigureFile(160mm,160mm){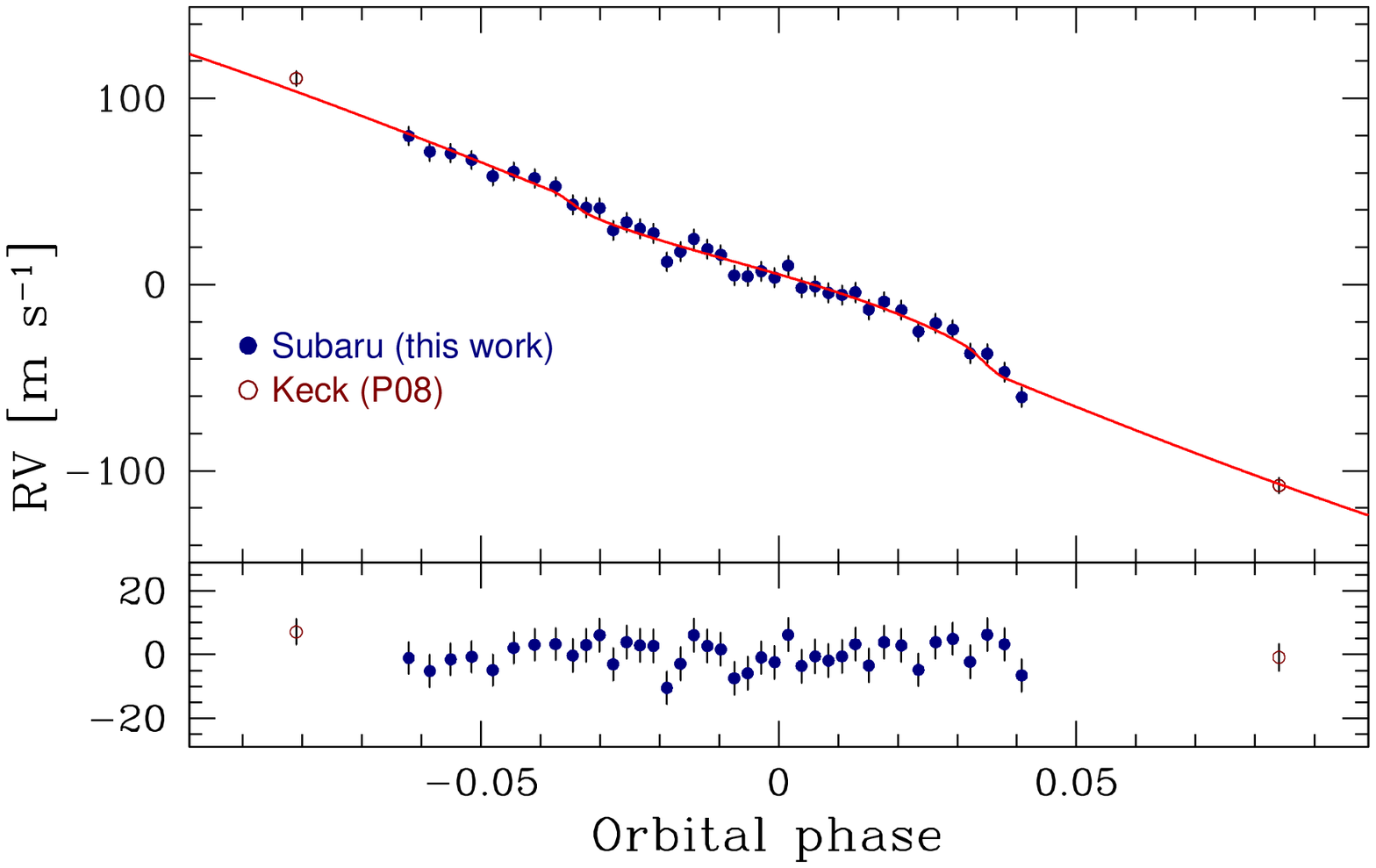}
 \end{center}
  \caption{
  Upper panels:
  Radial velocities and the best-fit curve of HAT-P-7
  as a function of orbital phase. The upper figure show the entire orbit
  and the lower figure do the zoom of transit phase.
  Bottom panels: Residuals from the best-fit curve.
  }
\end{figure*}
%%%%%%%%%%%%%%%%%%%%%%%%%%%%%%%%%%%%%%%%%%%%%%%%%%%%%%%%%%%%%%%%%%%%%%

\clearpage
%%%%%%%%%%%%%%%%%%%%%%%%%%%%%%%%%%%%%%%%%%%%%%%%%%%%%%%%%%%%%%%%%%%%%%
\begin{figure*}[pthb]
 \begin{center}
  \FigureFile(160mm,160mm){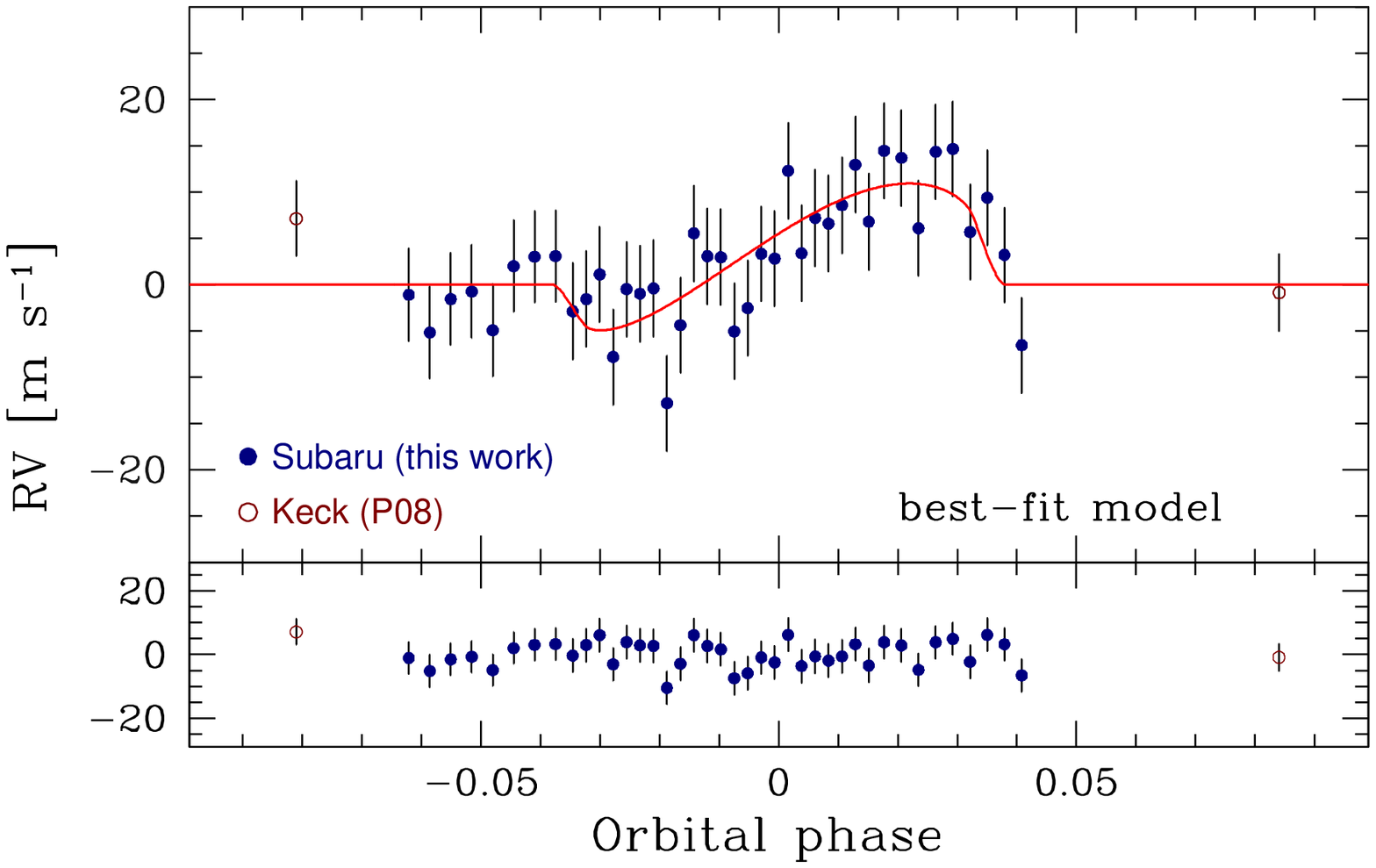}
  \FigureFile(160mm,160mm){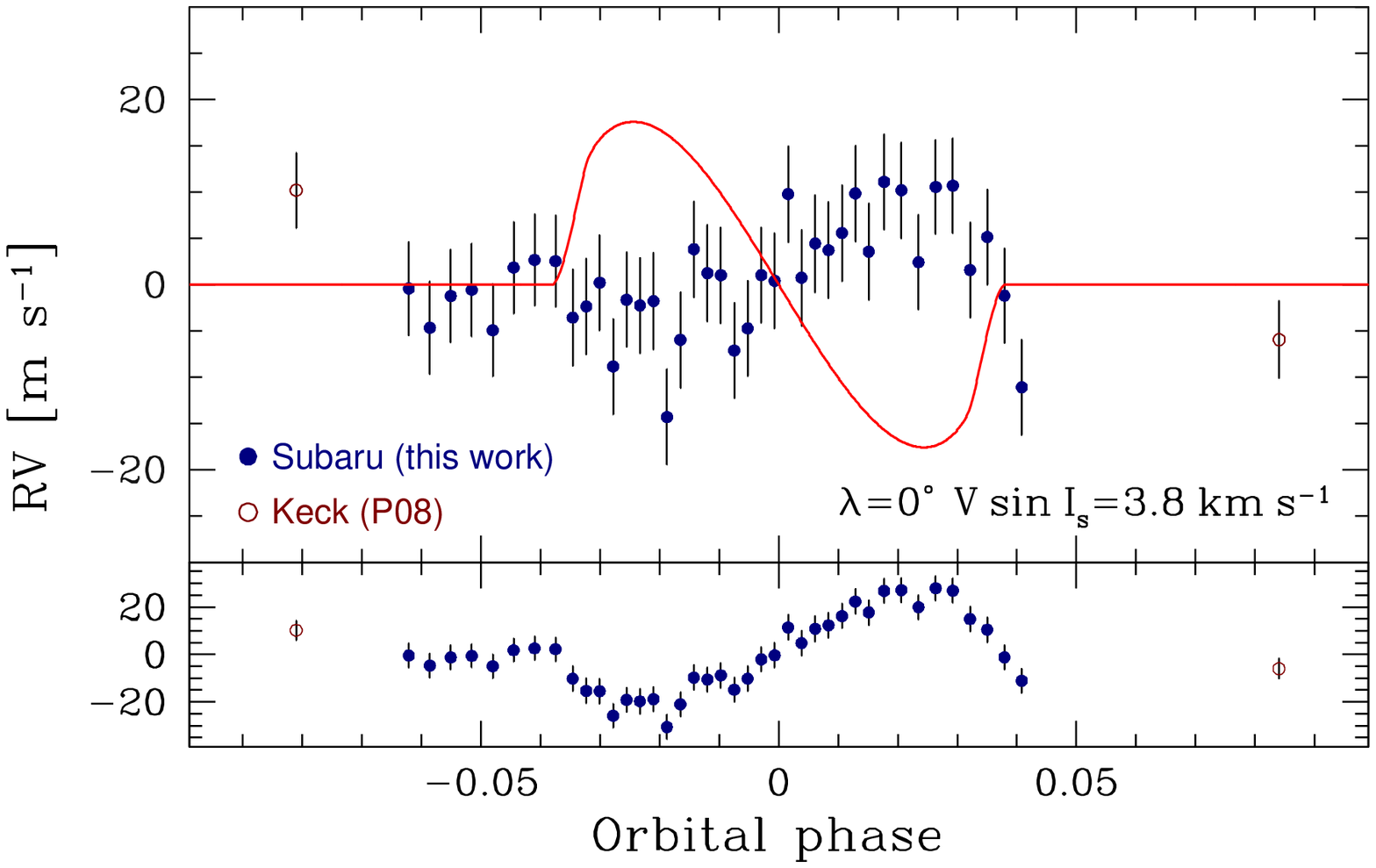}
 \end{center}
  \caption{
  Upper figure: The RM effect of HAT-P-7b. The upper panel shows
  difference radial velocities (namely, the Keplerian motion is
  subtracted from the original radial velocities).
  The solid line indicates the best-fit RM model.
  The lower panel plots residuals from the best-fit model.
  Lower figure: The case for $\lambda=0^{\circ}$ and
  $V \sin I_s = 3.8$~km~s$^{-1}$ model, for reference.
  }
\end{figure*}
%%%%%%%%%%%%%%%%%%%%%%%%%%%%%%%%%%%%%%%%%%%%%%%%%%%%%%%%%%%%%%%%%%%%%%

\clearpage
%%%%%%%%%%%%%%%%%%%%%%%%%%%%%%%%%%%%%%%%%%%%%%%%%%%%%%%%%%%%%%%%%%%%%%
\begin{figure*}[pthb]
 \begin{center}
  \FigureFile(120mm,120mm){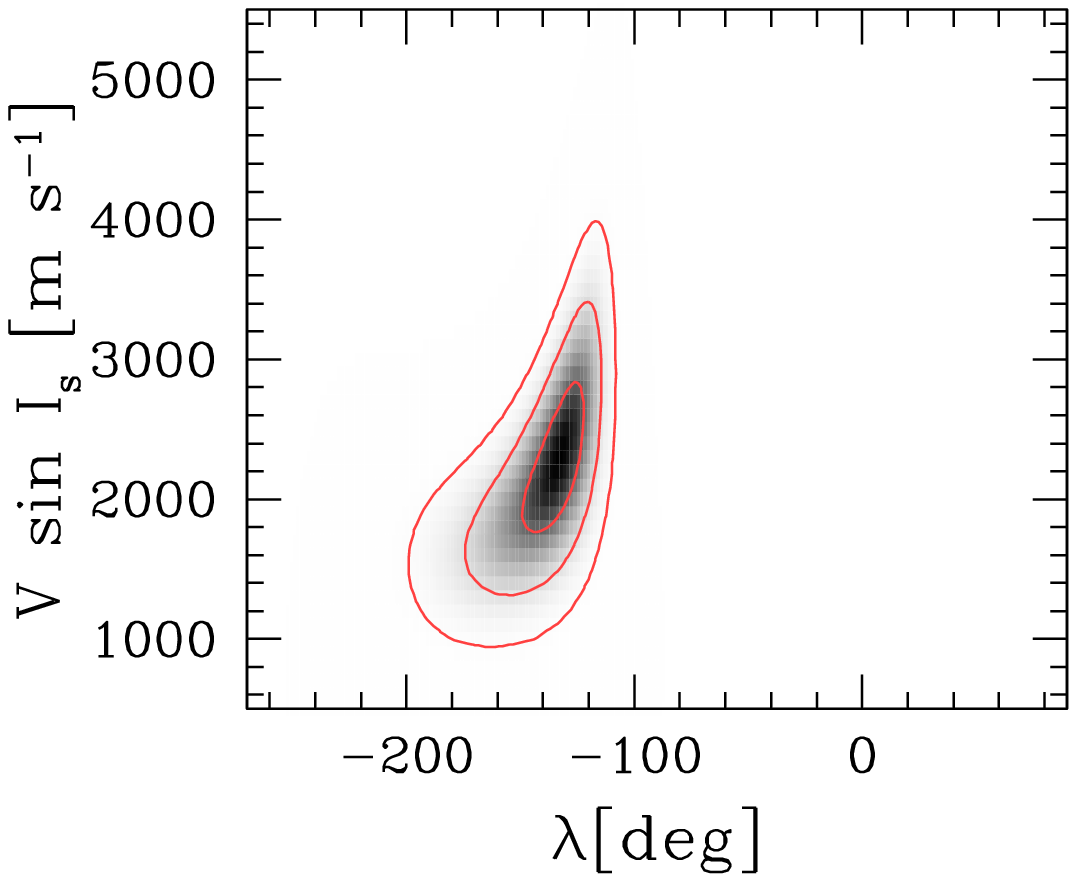}
  \FigureFile(120mm,120mm){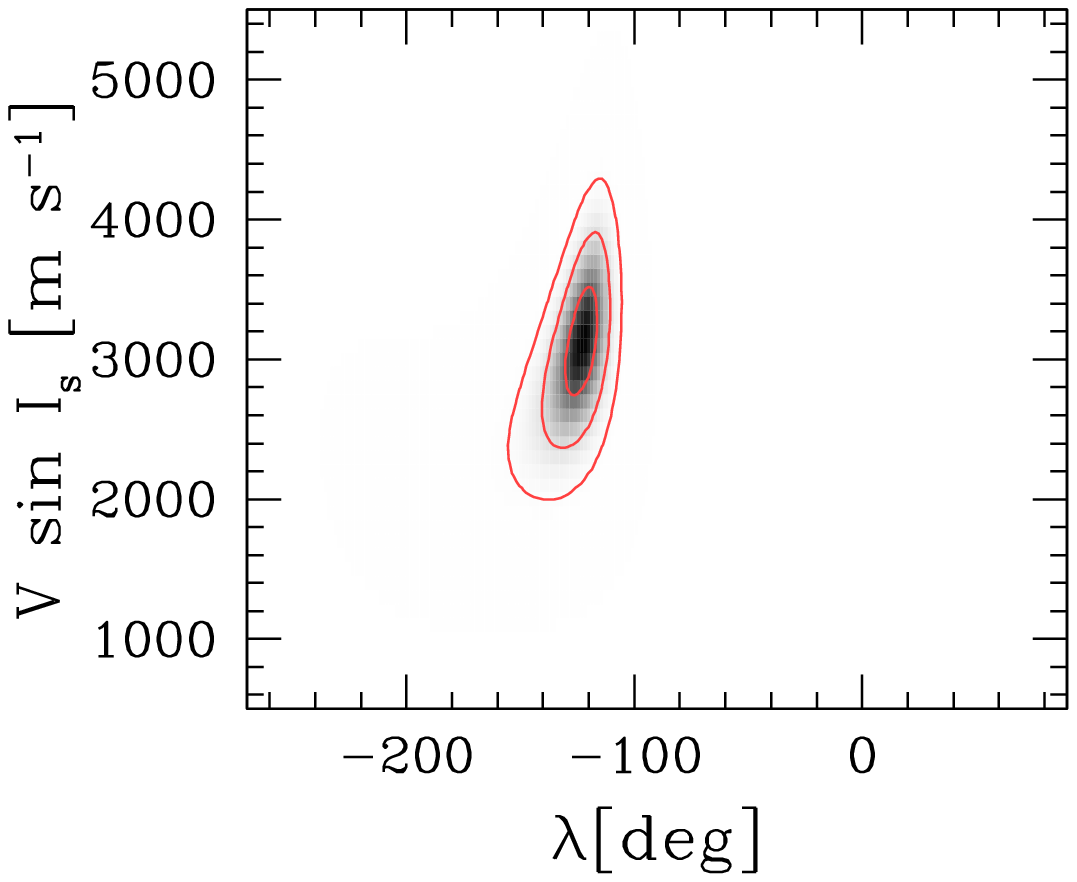}
 \end{center}
  \caption{
  Plots of ($\lambda,V \sin I_s$) contours of HAT-P-7 based on
  our Subaru dataset and the P08 Keck dataset without (upper) and
  with (lower) the \textit{a priori} constraint on $V \sin I_s$.
  The solid lines show $\Delta\chi^2$ = 1.0, 4.0, and 9.0
  (from inner to outer), respectively.
  }
\end{figure*}
%%%%%%%%%%%%%%%%%%%%%%%%%%%%%%%%%%%%%%%%%%%%%%%%%%%%%%%%%%%%%%%%%%%%%%

\clearpage
%%%%%%%%%%%%%%%%%%%%%%%%%%%%%%%%%%%%%%%%%%%%%%%%%%%%%%%%%%%%%%%%%%%%%%
\begin{longtable}[c]{lcc}
\caption{Radial velocities obtained with the Subaru/HDS.}
\hline
\endhead
\hline
\endfoot
\hline
Time [HJD]  & Value [m~s$^{-1}$] & Error [m~s$^{-1}$]\\
\hline
2454616.89606 &	65.07 &	5.02\\
2454616.90384 &	56.63 &	5.00\\
2454616.91160 &	55.84 &	4.98\\
2454616.91936 &	52.22 &	5.03\\
2454616.92712 &	43.55 &	5.00\\
2454616.93489 &	45.98 &	4.98\\
2454616.94264 &	42.45 &	4.94\\
2454616.95039 &	37.97 &	4.94\\
2454616.95678 &	28.22 &	5.20\\
2454616.96176 &	26.60 &	5.17\\
2454616.96674 &	26.32 &	5.15\\
2454616.97172 &	14.40 &	5.17\\
2454616.97670 &	18.77 &	5.11\\
2454616.98167 &	15.30 &	5.15\\
2454616.98666 &	12.88 &	5.21\\
2454616.99165 &	-2.53 &	5.18\\
2454616.99663 &	2.91 &	5.16\\
2454617.00161 &	9.84 &	5.17\\
2454617.00661 &	4.34 &	5.19\\
2454617.01159 &	1.23 &	5.17\\
2454617.01657 &	-9.81 &	5.16\\
2454617.02156 &	-10.32 &	5.16\\
2454617.02655 &	-7.47 &	5.13\\
2454617.03154 &	-11.00 &	5.15\\
2454617.03652 &	-4.53 &	5.17\\
2454617.04151 &	-16.47 &	5.16\\
2454617.04650 &	-15.68 &	5.23\\
2454617.05147 &	-19.28 &	5.16\\
2454617.05645 &	-20.31 &	5.21\\
2454617.06142 &	-18.94 &	5.17\\
2454617.06640 &	-28.10 &	5.21\\
2454617.07206 &	-23.85 &	5.16\\
2454617.07842 &	-28.44 &	5.18\\
2454617.08479 &	-39.88 &	5.13\\
2454617.09115 &	-35.41 &	5.09\\
2454617.09751 &	-38.90 &	5.13\\
2454617.10389 &	-51.68 &	5.15\\
2454617.11026 &	-51.74 &	5.16\\
2454617.11664 &	-61.71 &	5.12\\
2454617.12301 &	-75.20 &	5.14\\
\hline
\end{longtable}
%%%%%%%%%%%%%%%%%%%%%%%%%%%%%%%%%%%%%%%%%%%%%%%%%%%%%%%%%%%%%%%%%%%%%%

%%%%%%%%%%%%%%%%%%%%%%%%%%%%%%%%%%%%%%%%%%%%%%%%%%%%%%%%%%%%%%%%%%%%%%
\begin{table}[t]
\caption{Adopted stellar and planetary parameters.}
\begin{center}
\begin{tabular}{l|cc}
\hline
Parameter & Value & Source \\
\hline
$M_s$ [$M_{\odot}$] 
& $1.47$ & P08 \\
$R_s$ [$R_{\odot}$]
& $1.84$ & P08 \\
$R_p/R_s$ 
& $0.0763$ & P08 \\
$i$ [$^{\circ}$]
& $85.7$ & P08 \\
$a / R_s$
& $4.35$ & P08 \\
$u_1$
& $0.45$ & \citet{2004A&A...428.1001C} \\
$u_2$
& $0.31$ & \citet{2004A&A...428.1001C} \\
$T_c$ [HJD]
& $2453790.2593$ & P08 \\
$P$ [days]
& $2.2047299$ & P08 \\
\hline
\multicolumn{3}{l}{\hbox to 0pt{\parbox{80mm}{\footnotesize
}\hss}}
\end{tabular}
\end{center}
\end{table}
%%%%%%%%%%%%%%%%%%%%%%%%%%%%%%%%%%%%%%%%%%%%%%%%%%%%%%%%%%%%%%%%%%%%%%

%%%%%%%%%%%%%%%%%%%%%%%%%%%%%%%%%%%%%%%%%%%%%%%%%%%%%%%%%%%%%%%%%%%%%%
\begin{table}[t]
\caption{Best-fit values and uncertainties of the free parameters.}
\begin{center}
\begin{tabular}{l|cc|cc}
\hline
 & \multicolumn{2}{c|}{main-case}
 & \multicolumn{2}{c}{test-case} \\
Parameter & Value & Uncertainty & Value & Uncertainty \\
\hline
$K$ [m s$^{-1}$] 
& 212.6 & $\pm1.9$ 
& 213.3 & $\pm1.9$ \\
$V \sin I_s$ [km s$^{-1}$]\,$^{\rm{a}}$
& 2.3  & $^{+0.6}_{-0.5}$
& 3.1  & $\pm 0.4$ \\
$\lambda$ [$^{\circ}$]\,$^{\rm{a}}$
& -132.6  & $^{+10.5}_{-16.3}$
& -122.5  & $^{+6.4}_{-7.7}$\\
$v_1$ [m s$^{-1}$] 
& -14.7  & $\pm 1.6$
& -16.6  & $\pm 1.3$\\
rms (Subaru) [m s$^{-1}$] 
& 4.14  & 
& 4.32  & \\
$v_2$ [m s$^{-1}$] 
& -37.4  & $\pm 1.6$
& -37.5  & $\pm 1.6$\\
rms (Keck) [m s$^{-1}$] 
& 4.09  & 
& 4.09  & \\
\hline
\multicolumn{5}{l}{\hbox to 0pt{\parbox{100mm}{
\footnotemark[a]: Systematic errors are not included in the uncertainties
(see text).}\hss}}
\end{tabular}
\end{center}
\end{table}
%%%%%%%%%%%%%%%%%%%%%%%%%%%%%%%%%%%%%%%%%%%%%%%%%%%%%%%%%%%%%%%%%%%%%%


\begin{thebibliography}{}
\expandafter\ifx\csname natexlab\endcsname\relax\def\natexlab#1{#1}\fi

\bibitem[{{Bakos} {et~al.}(2007){Bakos}, {Kov{\'a}cs}, {Torres}, {Fischer},
  {Latham}, {Noyes}, {Sasselov}, {Mazeh}, {Shporer}, {Butler}, {Stefanik},
  {Fern{\'a}ndez}, {Sozzetti}, {P{\'a}l}, {Johnson}, {Marcy}, {Winn}, {Sip{\H
  o}cz}, {L{\'a}z{\'a}r}, {Papp}, \& {S{\'a}ri}}]{2007ApJ...670..826B}
{Bakos}, G.~{\'A}., {et~al.} 2007, \apj, 670, 826

\bibitem[{{Borucki} {et~al.}(2009){Borucki}, {Koch}, {Jenkins}, {Sasselov},
  {Gilliland}, {Batalha}, {Latham}, {Caldwell}, {Basri}, {Brown},
  {Christensen-Dalsgaard}, {Cochran}, {DeVore}, {Dunham}, {Dupree}, {Gautier},
  {Geary}, {Gould}, {Howell}, {Kjeldsen}, {Lissauer}, {Marcy}, {Meibom},
  {Morrison}, \& {Tarter}}]{2009Sci...325..709B}
{Borucki}, W.~J., {et~al.} 2009, Science, 325, 709

\bibitem[{{Butler} {et~al.}(1996){Butler}, {Marcy}, {Williams}, {McCarthy},
  {Dosanjh}, \& {Vogt}}]{1996PASP..108..500B}
{Butler}, R.~P., {Marcy}, G.~W., {Williams}, E., {McCarthy}, C., {Dosanjh}, P.,
  \& {Vogt}, S.~S. 1996, \pasp, 108, 500

\bibitem[{{Chatterjee} {et~al.}(2008){Chatterjee}, {Ford}, {Matsumura}, \&
  {Rasio}}]{2008ApJ...686..580C}
{Chatterjee}, S., {Ford}, E.~B., {Matsumura}, S., \& {Rasio}, F.~A. 2008, \apj,
  686, 580

\bibitem[{{Claret}(2004)}]{2004A&A...428.1001C}
{Claret}, A. 2004, \aap, 428, 1001

\bibitem[{{Coelho} {et~al.}(2005){Coelho}, {Barbuy}, {Mel{\'e}ndez},
  {Schiavon}, \& {Castilho}}]{2005A&A...443..735C}
{Coelho}, P., {Barbuy}, B., {Mel{\'e}ndez}, J., {Schiavon}, R.~P., \&
  {Castilho}, B.~V. 2005, \aap, 443, 735

\bibitem[{{Fabrycky} \& {Tremaine}(2007)}]{2007ApJ...669.1298F}
{Fabrycky}, D., \& {Tremaine}, S. 2007, \apj, 669, 1298

\bibitem[{{Fabrycky} \& {Winn}(2009)}]{2009ApJ...696.1230F}
{Fabrycky}, D.~C., \& {Winn}, J.~N. 2009, \apj, 696, 1230

\bibitem[{{Gaudi} \& {Winn}(2007)}]{2007ApJ...655..550G}
{Gaudi}, B.~S., \& {Winn}, J.~N. 2007, \apj, 655, 550

\bibitem[{{Gray}(2005)}]{2005oasp.book.....G}
{Gray}, D.~F. 2005, {The Observation and Analysis of Stellar Photospheres}

\bibitem[{{H{\'e}brard} {et~al.}(2008){H{\'e}brard}, {Bouchy}, {Pont},
  {Loeillet}, {Rabus}, {Bonfils}, {Moutou}, {Boisse}, {Delfosse}, {Desort},
  {Eggenberger}, {Ehrenreich}, {Forveille}, {Lagrange}, {Lovis}, {Mayor},
  {Pepe}, {Perrier}, {Queloz}, {Santos}, {S{\'e}gransan}, {Udry}, \&
  {Vidal-Madjar}}]{2008A&A...488..763H}
{H{\'e}brard}, G., {et~al.} 2008, \aap, 488, 763

\bibitem[{{Henry} \& {Winn}(2008)}]{2008AJ....135...68H}
{Henry}, G.~W., \& {Winn}, J.~N. 2008, \aj, 135, 68

\bibitem[{{Ida} \& {Lin}(2004)}]{2004ApJ...616..567I}
{Ida}, S., \& {Lin}, D.~N.~C. 2004, \apj, 616, 567

\bibitem[{{Johnson} {et~al.}(2009){Johnson}, {Winn}, {Albrecht}, {Howard},
  {Marcy}, \& {Gazak}}]{2009arXiv0907.5204J}
{Johnson}, J.~A., {Winn}, J.~N., {Albrecht}, S., {Howard}, A.~W., {Marcy},
  G.~W., \& {Gazak}, J.~Z. 2009, arXiv:0907.5204

\bibitem[{{Lin} {et~al.}(1996){Lin}, {Bodenheimer}, \&
  {Richardson}}]{1996Natur.380..606L}
{Lin}, D.~N.~C., {Bodenheimer}, P., \& {Richardson}, D.~C. 1996, \nat, 380, 606

\bibitem[{{Lin} \& {Papaloizou}(1985)}]{1985prpl.conf..981L}
{Lin}, D.~N.~C., \& {Papaloizou}, J. 1985, in Protostars and Planets II, ed.
  D.~C. {Black} \& M.~S. {Matthews}, 981--1072

\bibitem[{{Marzari} \& {Weidenschilling}(2002)}]{2002Icar..156..570M}
{Marzari}, F., \& {Weidenschilling}, S.~J. 2002, Icarus, 156, 570

\bibitem[{{McLaughlin}(1924)}]{1924ApJ....60...22M}
{McLaughlin}, D.~B. 1924, \apj, 60, 22

\bibitem[{{Moutou} {et~al.}(2009){Moutou}, {H{\'e}brard}, {Bouchy},
  {Eggenberger}, {Boisse}, {Bonfils}, {Gravallon}, {Ehrenreich}, {Forveille},
  {Delfosse}, {Desort}, {Lagrange}, {Lovis}, {Mayor}, {Pepe}, {Perrier},
  {Pont}, {Queloz}, {Santos}, {S{\'e}gransan}, {Udry}, \&
  {Vidal-Madjar}}]{2009A&A...498L...5M}
{Moutou}, C., {et~al.} 2009, \aap, 498, L5

\bibitem[{{Nagasawa} {et~al.}(2008){Nagasawa}, {Ida}, \&
  {Bessho}}]{2008ApJ...678..498N}
{Nagasawa}, M., {Ida}, S., \& {Bessho}, T. 2008, \apj, 678, 498

\bibitem[{{Narita} {et~al.}(2008){Narita}, {Sato}, {Ohshima}, \&
  {Winn}}]{2008PASJ...60L...1N}
{Narita}, N., {Sato}, B., {Ohshima}, O., \& {Winn}, J.~N. 2008, \pasj, 60, L1+

\bibitem[{{Narita} {et~al.}(2007){Narita}, {Enya}, {Sato}, {Ohta}, {Winn},
  {Suto}, {Taruya}, {Turner}, {Aoki}, {Tamura}, {Yamada}, \&
  {Yoshii}}]{2007PASJ...59..763N}
{Narita}, N., {et~al.} 2007, \pasj, 59, 763

\bibitem[{{Narita} {et~al.}(2009){Narita}, {Hirano}, {Sato}, {Winn}, {Suto},
  {Turner}, {Aoki}, {Tamura}, \& {Yamada}}]{2009arXiv0905.4727N}
{Narita}, N., {et~al.} 2009, arXiv:0905.4727

\bibitem[{{Noguchi} {et~al.}(2002){Noguchi}, {Aoki}, {Kawanomoto}, {Ando},
  {Honda}, {Izumiura}, {Kambe}, {Okita}, {Sadakane}, {Sato}, {Tajitsu},
  {Takada-Hidai}, {Tanaka}, {Watanabe}, \& {Yoshida}}]{2002PASJ...54..855N}
{Noguchi}, K., {et~al.} 2002, \pasj, 54, 855

\bibitem[{{Ohta} {et~al.}(2005){Ohta}, {Taruya}, \&
  {Suto}}]{2005ApJ...622.1118O}
{Ohta}, Y., {Taruya}, A., \& {Suto}, Y. 2005, \apj, 622, 1118

\bibitem[{{P{\'a}l} {et~al.}(2008){P{\'a}l}, {Bakos}, {Torres}, {Noyes},
  {Latham}, {Kov{\'a}cs}, {Marcy}, {Fischer}, {Butler}, {Sasselov}, {Sip{\H
  o}cz}, {Esquerdo}, {Kov{\'a}cs}, {Stefanik}, {L{\'a}z{\'a}r}, {Papp}, \&
  {S{\'a}ri}}]{2008ApJ...680.1450P}
{P{\'a}l}, A., {et~al.} 2008, \apj, 680, 1450 (P08)

\bibitem[{{Pont} {et~al.}(2009){Pont}, {H{\'e}brard}, {Irwin}, {Bouchy},
  {Moutou}, {Ehrenreich}, {Guillot}, {Aigrain}, {Bonfils}, {Berta}, {Boisse},
  {Burke}, {Charbonneau}, {Delfosse}, {Desort}, {Eggenberger}, {Forveille},
  {Lagrange}, {Lovis}, {Nutzman}, {Pepe}, {Perrier}, {Queloz}, {Santos},
  {S{\'e}gransan}, {Udry}, \& {Vidal-Madjar}}]{2009A&A...502..695P}
{Pont}, F., {et~al.} 2009, \aap, 502, 695

\bibitem[{{Press} {et~al.}(1992){Press}, {Teukolsky}, {Vetterling}, \&
  {Flannery}}]{1992nrca.book.....P}
{Press}, W.~H., {Teukolsky}, S.~A., {Vetterling}, W.~T., \& {Flannery}, B.~P.
  1992, {Numerical recipes in C. The art of scientific computing} (Cambridge:
  University Press, |c1992, 2nd ed.)

\bibitem[{{Rasio} \& {Ford}(1996)}]{1996Sci...274..954R}
{Rasio}, F.~A., \& {Ford}, E.~B. 1996, Science, 274, 954

\bibitem[{{Rossiter}(1924)}]{1924ApJ....60...15R}
{Rossiter}, R.~A. 1924, \apj, 60, 15

\bibitem[{{Sato} {et~al.}(2002){Sato}, {Kambe}, {Takeda}, {Izumiura}, \&
  {Ando}}]{2002PASJ...54..873S}
{Sato}, B., {Kambe}, E., {Takeda}, Y., {Izumiura}, H., \& {Ando}, H. 2002,
  \pasj, 54, 873

\bibitem[{{Sozzetti} {et~al.}(2009){Sozzetti}, {Torres}, {Charbonneau}, {Winn},
  {Korzennik}, {Holman}, {Latham}, {Laird}, {Fernandez}, {O'Donovan},
  {Mandushev}, {Dunham}, {Everett}, {Esquerdo}, {Rabus}, {Belmonte}, {Deeg},
  {Brown}, {Hidas}, \& {Baliber}}]{2009ApJ...691.1145S}
{Sozzetti}, A., {et~al.} 2009, \apj, 691, 1145

\bibitem[{{Takeda} \& {Rasio}(2005)}]{2005ApJ...627.1001T}
{Takeda}, G., \& {Rasio}, F.~A. 2005, \apj, 627, 1001

\bibitem[{{Winn} {et~al.}(2005){Winn}, {Noyes}, {Holman}, {Charbonneau},
  {Ohta}, {Taruya}, {Suto}, {Narita}, {Turner}, {Johnson}, {Marcy}, {Butler},
  \& {Vogt}}]{2005ApJ...631.1215W}
{Winn}, J.~N., {et~al.} 2005, \apj, 631, 1215

\bibitem[{{Winn} {et~al.}(2007){Winn}, {Johnson}, {Peek}, {Marcy}, {Bakos},
  {Enya}, {Narita}, {Suto}, {Turner}, \& {Vogt}}]{2007ApJ...665L.167W}
{Winn}, J.~N., {et~al.} 2007, \apjl, 665, L167

\bibitem[{{Winn} {et~al.}(2009{\natexlab{a}}){Winn}, {Johnson}, {Fabrycky},
  {Howard}, {Marcy}, {Narita}, {Crossfield}, {Suto}, {Turner}, {Esquerdo}, \&
  {Holman}}]{2009ApJ...700..302W}
{Winn}, J.~N., {et~al.} 2009{\natexlab{a}}, \apj, 700, 302

\bibitem[{{Winn} {et~al.}(2009{\natexlab{b}}){Winn}, {Howard}, {Johnson},
  {Marcy}, {Gazak}, {Starkey}, {Ford}, {Colon}, {Reyes}, {Nortmann},
  {Dreizler}, {Odewahn}, {Welsh}, {Kadakia}, {Vanderbei}, {Adams}, {Lockhart},
  {Crossfield}, {Valenti}, {Dantowitz}, \& {Carter}}]{2009arXiv0907.5205W}
{Winn}, J.~N., {et~al.} 2009{\natexlab{b}}, arXiv:0907.5205

\bibitem[{{Winn} {et~al.}(2009{\natexlab{c}}){Winn}, {Johnson}, {Albrecht},
  {Howard}, {Marcy}, {Crossfield}, \& {Holman}}]{2009arXiv0908.1672W}
{Winn}, J.~N., {Johnson}, J.~A., {Albrecht}, S., {Howard}, A.~W., {Marcy},
  G.~W., {Crossfield}, I.~J., \& {Holman}, M.~J. 2009{\natexlab{c}},
arXiv:0908.1672

\bibitem[{{Wu} \& {Murray}(2003)}]{2003ApJ...589..605W}
{Wu}, Y., \& {Murray}, N. 2003, \apj, 589, 605

\bibitem[{{Wu} {et~al.}(2007){Wu}, {Murray}, \&
  {Ramsahai}}]{2007ApJ...670..820W}
{Wu}, Y., {Murray}, N.~W., \& {Ramsahai}, J.~M. 2007, \apj, 670, 820

\end{thebibliography}
\end{document}